# Temperature Prediction for Stored Grain: A Multi-model Fusion Approach Based on Machine Learning

Donghao Chen
Key Laboratory of Grain Information Processing and Control
Henan University of Technology
Ministry of Education
Zhengzhou, China
chendonghao@haut.edu.cn

* Binkun Liu
Henan Engineering Laboratory of Grain Condition Intelligent Detection and Application
Henan University of Technology
Zhengzhou, China
*Correspondence: bkliu@haut.edu.cn

*Abstract*—Temperature fluctuations significantly affect microorganism growth and pest activities in grain pile, precise monitoring and forecasting temperature of stored grain are essential for maintaining the quality and safety of grain storage. This paper proposes a multi-model fusion approach to predict grain temperature using historical temperature data of stored grain and meteorological data from the region. Firstly, four distinct machine learning models, namely Adaboost, decision tree, extra trees, and random forest, are fine-tuned through parameter optimization to enhance their predictive capabilities respectively; Subsequently, these optimized models are fused to form different ensemble models, which are compared for prediction accuracy to obtain the optimal fusion model. In essence, the fusion process integrates the predictions of each individual model as new feature inputs into the fusion models. Furthermore, random forest is utilized to identify the key factors influencing grain temperature, providing insights into the importance of different influencing factors. The experimental results demonstrate that the proposed fusion models can achieve higher prediction accuracy and robustness compared with the single-model prediction methods. Additionally, the analysis of feature importance also offers empirical evidence for understanding the factors influencing grain temperature.

## I. INTRODUCTION

As the world's population increases, food demand is projected to increase by 70% by 2050 due to increased population and per capita consumption. To meet this demand, annual cereal production will need to increase to over 3 billion tons by 2050 [1]. Five UN specialized agencies noted in their report that since 2019, with the recurrence of COVID-19 pandemic and the turbulence of events such as climatic shocks and the conflict in Ukraine, the world has added more hungry population of more than 122 million. If the situation is allowed to develop, the world will not be able to achieve the sustainable development goal of eradicating hunger by 2030 as scheduled [2]. Therefore, the importance of food storage cannot be ignored.

Grain temperature is an important indicator of grain pile conditions. Therefore, effective monitoring and prediction of the temperature and humidity of grain piles are key measures to prevent grain damage. Currently, all levels of grain depots in China have adopted temperature sensor arrays and grain information management systems to monitor and analyze the temperature and humidity data within the grain piles in real time to determine whether the grain is safe or not. However, the functions of these systems are limited, they cannot accurately reflect the correctness and changing trend of the grain temperature, and they still need to rely on the custodian's experience to make judgment. If advanced technologies and methods can be utilized to make reasonable predictions of future temperatures in grain piles and take timely protective measures accordingly, it will be of great significance to improve food security.

In the early research field of grain temperature prediction, the prediction methods were mainly based on traditional mathematical methods. For example, as early as 1965, Boyce proposed that in the process of grain drying, through the analysis of the physical properties of the grain itself, so as to establish a set of equations for the relationship between its temperature changes with the grain to make predictions on the changes in stored grain temperature during the drying process [3]. Jia et al. employed the finite element analysis method to establish a two-dimensional nonlinear heat transfer model, which was used to predict the temperature distribution of the vertical silo [4]. Yan et al. proposed a storage temperature detection method based on acoustic tomography to model the relationship between stored grain temperature and sound travel time [5]. A model utilizing finite element method was proposed for simulating transient heat, mass, momentum, and species transfer in the stored grain ecosystem in three dimensions [6]. However, due to the complexity of the factors affecting the change of grain temperature, it is challenging to predict the stored grain temperature based on mathematical modeling. After the 21st century, with the rapid development of artificial intelligence, methods such as machine learning began to be widely used. For example, Yang et al. used machine learning algorithms to estimate humidity and mold conditions in stored wheat [7, 8]. Duan et al. used Support Vector Machine (SVM) and Adaptive Boosting (AdaBoost) to predict grain temperature [9, 10], and Wang et al. [11] combined thermodynamic model and spatial stochastic process theory to propose a simulation learning method for predicting the internal temperature of grain piles in a tall bungalow silo. Li et al. used Autoregressive Integrated Moving Average Model (ARIMA) and Holt-Winters model to predict stored grain temperature, and both of them achieved good results [12]. However, the robustness and the prediction performance of the above research models could be further improved. Moreover, the importance of various features affecting grain temperatureare is rarely studied.

To solve the above problems, in this paper, a multi-model fusion approach is developed to predict the stored grain temperature. Firstly, this study utilizes different machine learning models, including four models, Adaboost, decision tree, extra trees, and random forest. Then, different model parameters are set to train the models with grain temperature data so as to get a single best prediction model respectively. Then the above four machine learning models are fused to train the 11 superior fusion models, and compared with the single best model, according to the calculation of the relevant model evaluation criteria, it is clear that the method of multi-model fusion has higher accuracy and can better predict the stored grain temperature. At the same time, this paper uses the random forest feature importance analysis experiment to derive the important factors affecting the temperature of the stored grain. Compared with the traditional single model algorithms, the multi-model fusion approach based on machine learning may provide helpful references for the temperature prediction of stored grain.

## II. METHOD

### A. Data preparation

The experimental dataset used in this study was taken from a large grain depot in Hunan Province, China, including wheat temperature data from eight granaries and corresponding meteorological data from the China Meteorological Data Network. The dataset was collected from January 1, 2020 to June 8, 2021, and each piece of data includes five parameters: warehouse temperature, warehouse humidity, air temperature, air humidity, and average grain temperature of the whole warehouse. 140 temperature sensors are deployed in the granary, in which 7 rows of sensors are arranged in the x-axis direction, 5 rows of sensors are arranged in the y-axis direction, and 4 layers of sensors are arranged along the z-axis. In this experiment, the average value of temperature data measured by each layer of sensors is taken as the average grain temperature of the whole granary, where the four parameters of warehouse temperature, warehouse humidity, air temperature, and air humidity are used as the feature inputs. In addition, the dataset is partitioned into two parts, namely the training set and the test set, with a ratio of 7:3.

### B. Machine learning models

*1) Adaboost:* Adaboost [13] is a powerful integrated learning algorithm that enhances a weak learner with a prediction accuracy that is only slightly better than a random guess into a strong learner with a high prediction accuracy. It does this by iteratively training a series of weak classifiers and adjusting the sample weights according to their performance, eventually combining these weak classifiers into a strong classifier. The features of Adaboost include an upper bound on the error rate that decreases with the number of iterations, and it is less prone to overfitting. The algorithm flow of Adaboost is as follows:

**Step 1:** Given dataset $(X,Y) = \{(x_1, y_1),(x_i, y_i),...,(x_N, y_N)\}$, where $x_i \in X \subseteq R^n$ is the sample space, $y_i \in Y = \{1,-1\}$ denotes the sample labels. Initialize the weight distribution of the training data. Each training sample is assigned the same weight at the very beginning: $w_{1i} = 1/N$, such that the initial weight distribution of the training sample set $D_1$:

$$D_1 = (w_{11}, w_{12},...w_{1N}) = (1/N,...,1/N) \quad (1)$$

**Step 2:** Perform iterations $t = 1,...,T$

Firstly, select a weak classifier $h_t$ with the lowest current error rate as the *t*th basic classifier $H_t$, and compute the weak classifier $h_t : X \to \{-1,1\}$, the error rate of this weak classifier on the distribution $D_t$ is:

$$e_t = P(H_t(x_i) \neq y_i) = \sum_{i=1}^{N} w_{ti} I(H_t(x_i) \neq y_i) \quad (2)$$

From equation (2), the error rate $e_t$ of $H_t(x)$ on the training dataset is the sum of the weights of the samples misclassified by $H_t(x)$.

Then, calculate the weight of this weak classifier in the final classifier (weak classifier weight is denoted by $\boldsymbol{\alpha}$):

$$\boldsymbol{\alpha}_t = \frac{1}{2} \ln(\frac{1-e_t}{e_t}) \quad (3)$$

Finally, update the weight distribution $D_{t+1}$ of the training samples:

$$D_{t+1}(i) = D_t(i) \exp(-\boldsymbol{\alpha}_t y_i H_t(x_i))/Z_t \quad (4)$$

where $Z_t$ is a normalization constant: $Z_t = 2\sqrt{e_t(1-e_t)}$

**Step 3:** Each weak classifier is combined according to the weak classifier weight $\boldsymbol{\alpha}_t$ i.e.

$$f(x) = \sum_{i=1}^{T} \boldsymbol{\alpha}_t H_t(x) \quad (5)$$

By the action of the sign function sign, a strong classifier is obtained as:

$$H_{final} = sign(f(x)) = sign(\sum_{i=1}^{T} \boldsymbol{\alpha}_t H_t(x)) \quad (6)$$

*2) Decision tree:* Decision tree is a model for decision making based on a tree structure. It categorizes the data set through multiple conditional discriminative processes and finally obtains the desired results. Decision tree starts from the root node and is gradually divided into internal nodes and leaf nodes. The internal nodes represent the feature attribute tests and the leaf nodes represent the decision results. Each path from the root node to the leaf nodes forms a classification rule that recursively obtains the conclusion. Fig. 1 shows the decision tree model, a detailed explanation is given in [14].

*3) Random forest:* Random forest [15] uses the idea of Bagging, given a training set $X = x_1,...,x_n$ and a target $Y = y_1,...,y_n$, the bagging method repetitively ($B$ times) samples from the training set with putbacks, and then trains a

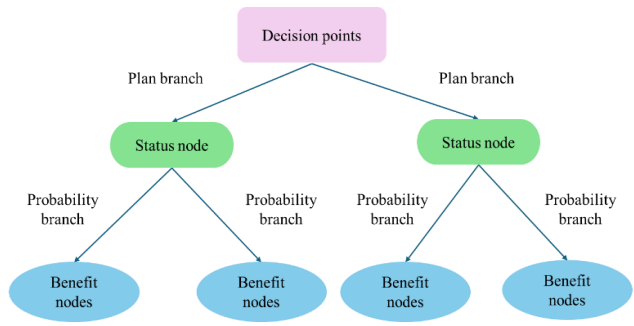

Figure 1. Decision tree model

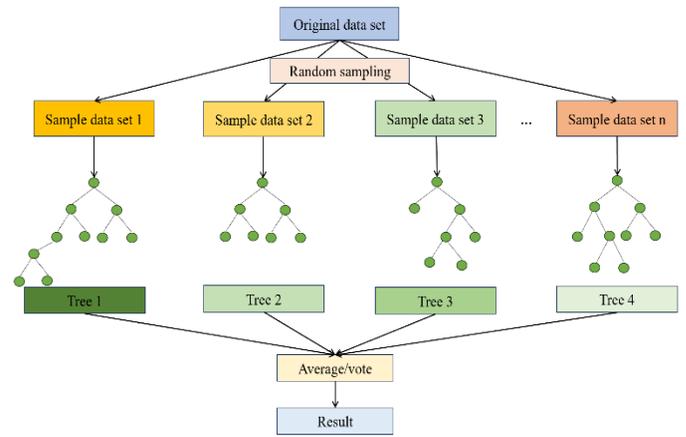

Figure 2. Random forest model

tree model on these samples, for *b=1,..., B:*

i) Sample, with replacement, *n* training examples from $X$, $Y$, call these $X_b$, $Y_b$.

ii) Train a classification or regression tree $f_b$ on $X_b$, $Y_b$.

Random forests take decision trees as the basic unit, by integrating a large number of decision trees, it constitutes a random forest. As shown in Fig. 2, the random forest model randomly samples the original data set, constituting n different sample data sets, and then builds *n* different decision tree models based on these data sets, and finally obtains the final results based on the average (for regression models) or voting (for classification models) of these decision tree models.

Moreover, the random forest can naturally be used to rank the importance of variables in a regression or classification problem. In this article, the feature importance of random forest in sklearn is employed to analyze the feature importance of the factors affecting the grain temperature and came up with the main factors affecting the grain temperature.

*4) Extra trees:* Extremely Randomized Trees (also known as Extra Trees) is an integrated learning algorithm for decision trees. It is similar to Random forest in that it consists of many decision trees, but Extremely Randomized Trees introduces more randomness in the construction of the decision trees: first, each tree has been trained using the entire learning sample, and second, the top-down division is randomized. Instead of calculating the optimal division point for each feature (e.g., based on information entropy or Gini impurity), it chooses the division point randomly. The value is selected uniformly and randomly from the empirical range of features. Among all the randomized division points, the one with the highest score is chosen as the division point of the node [16].

*C. Multi-model fusion*

In this paper, a novel model fusion strategy is proposed. Initially, the original dataset is fed into four distinct models for training. After consistent parameterization, the optimal model is identified based on performance metrics. Subsequently, predictions for both the training set and the test set are generated using this optimal model. These predicted values from the four models are then aggregated into a new set of features.

The crux of this approach lies in leveraging these aggregated features as input for training and fine-tuning a random forest model. By doing so, eleven unique fusion models are created, each capitalizing on the strengths of the initial models while benefiting from the enriched feature space. This fusion approach aims to enhance robustness and accuracy, making it a valuable addition to predictive modeling.

This experiment is based on the jupyter notebook platform utilizing the sklearn library within the python3 environment. Fig. 3 illustrates the model fusion process of this experiment with the Adaboost-extra trees-decision tree-random forest fusion model, and the following are the specific steps for model fusion:

**Step 1:** Import the dataset and utilize the train_test_split function from the sklearn library to divide it into training and test sets, maintaining a 7:3 ratio. Then set the random_state parameter to 0 to ensure consistent data splits across the four models.

**Step 2:** Train four regression models: Adaboost, random forest, Extra trees, and Decision tree. The performance of Adaboost, Random Forest, and Extra Trees primarily depends on the n_estimators parameter, while Decision tree relies on parameters such as criterion, min_samples_split, and min_samples_leaf. Then iteratively adjust model parameters to obtain the optimal configuration for each model. Based on testing, the random forest model emerged as the best performer.

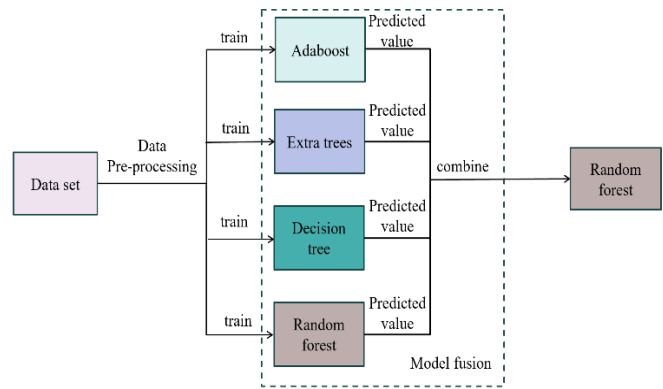

Figure 3. Fusion modeling process

TABLE I. PREDICTIVE PERFORMANCE OF DIFFERENT MODELS

| Model | n_estimators | MSE | $R^2$ |
|---|---|---|---|
| Adaboost | 50 | 0.1159 | 0.9825 |
| Decision tree | - | 0.1124 | 0.9830 |
| Extra trees | 84 | 0.1093 | 0.9835 |
| Random forest | 225 | 0.0795 | 0.9850 |
| Adaboost-decision tree | 244 | 0.0927 | 0.9859 |
| Decision tree-random forest | 2 | 0.0892 | 0.9865 |
| Adaboost-random forest | 2 | 0.0812 | 0.9877 |
| Decision tree-extra trees | 6 | 0.0783 | 0.9882 |
| Decision tree-extra trees-random forest | 2 | 0.0708 | 0.9893 |
| Adaboost-decision tree-random forest | 2 | 0.0688 | 0.9896 |
| Adaboost-extra trees | 16 | 0.0636 | 0.9904 |
| Adaboost-decision tree-Extra trees | 44 | 0.0627 | 0.9905 |
| Extratrees-random forest | 1 | 0.0603 | 0.9909 |
| Adaboost-extra trees-random forest | 2 | 0.0524 | 0.9921 |
| Adaboost-decision tree-extra trees-random forest | 8 | 0.0505 | 0.9924 |

**Step 3:** On the basis of obtaining the above single optimal model, generate predictions for both the training and test sets using the optimal random forest model. Then combine the predicted values from the four models to create a new set of features and utilize the aggregated features as input for training and fine-tuning an enhanced random forest model. Then repeat this process to create eleven distinct fusion models.

**Step 4:** Assess the performance of the obtained models using Mean Squared Error (MSE) and R-squared $R^2$. A smaller MSE indicates better predictive accuracy, while an $R^2$ value closer to 1 signifies superior data fitting.

## III. RESULTS AND DISCUSSION

### A. Prediction performance

In this experiment, by training four models Adaboost, Decision tree, Extra trees, random forest by traversing the method of tuning parameter and fusing the four models, the prediction performance results of 15 different models can be obtained shown in Table I. The n_estimators in the table are the main parameters of the regressors for each model to make the final prediction, among which the Decision tree model does not contain this parameter, and the fusion model ultimately uses the random forest regressor to learn the corresponding weights and make the prediction. From the above results, it can be seen that in the single model, the MSE of Random forest is the smallest and $R^2$ is the largest, which indicates that this model has the best prediction performance and the best fitting ability in the single model, while in the fusion model, the Adaboost-decision tree-extra trees-random forest model has the best performance, and its MSE can be as low as 0.0505, and the prediction performance is significantly better than the single model.

Furthermore, upon examining the outcomes, it's observed that in certain instances, the predictive accuracy of the standalone random forest model surpasses that of the combined fusion model. This could be attributed to the difficulty in integrating features learned by some models. However, generally speaking, the overall predictive efficiency and adaptability of the fusion model outperforms those of any single model. This superiority can be understood through two main perspectives. First, different models have different characteristics, such as Adaboost is excellent in dealing with high-dimensional data and complex classification problems, but it is sensitive to noisy data, by combining Adaboost with other models, its sensitivity to noisy data can be reduced and the stability of the overall model can be improved. The Decision tree can handle nonlinear relationships and feature interactions, but are insensitive to outliers, easy to overfit, and have limited generalization ability. Random forest can reduce the risk of overfitting and improve the stability and accuracy of the model. By combining random forest with other models, the diversity of the model can be further improved, and the prediction error can be reduced. Extra trees is similar to Random forest, but more random in constructing the tree, which can reduce the risk of overfitting and improve the generalization ability of the model. Therefore, the use of model fusion can comprehensively utilize their respective advantages and make up for their respective deficiencies, thus improving the overall model prediction effect. Second, the final choice of random forest regressor to learn the corresponding weights to do prediction has better prediction.

### B. Feature importance

In addition to obtaining the prediction performance results of the above model fusion, this paper also analyzed the four main features affecting grain temperature: warehouse temperature, warehouse humidity, air temperature, and air humidity using the feature importance of Random forest in the sklearn library.

The results were visualized as shown in Fig. 4, it can be observed that the impact weights of the four crucial factors affecting grain temperature follow this order: warehouse temperature, air temperature, warehouse humidity, and air humidity. Among these features, warehouse temperature exerts the most significant influence on grain temperature. In terms of specific analysis, warehouse temperature directly affects heat conduction and evaporation in grain piles, elevated warehouse temperatures may lead to spoilage or mold growth. Air temperature also plays a substantial role, external temperature conditions affect the internal warehouse temperature, if external temperatures are excessively high, the warehouse temperature rises, impacting grain temperature. Warehouse humidity is another critical factor for grain preservation, excessive humidity can cause grain to absorb moisture, making them susceptible to mold. Therefore, maintaining appropriate warehouse humidity is equally crucial. Air humidity, similar to warehouse humidity, is influenced more by external environmental conditions. Considering all these factors, the results obtained from the experiment are reasonable.

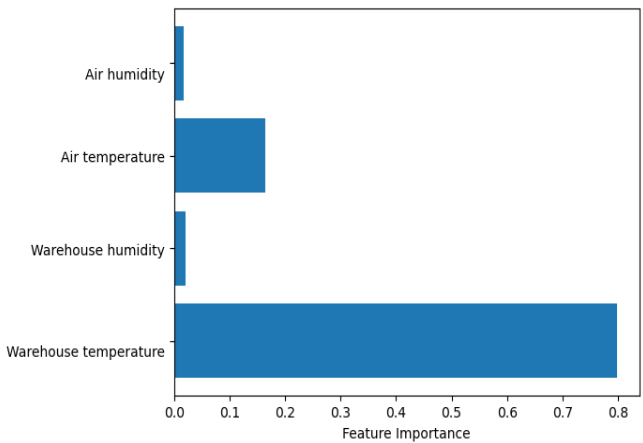

Figure 4. The importance of various features affecting grain temperature

## IV. Conclusion

This paper proposes a multi-model fusion approach based on machine learning to predict the temperature of stored grain. In this paper, four models are firstly utilized, namely, Adaboost, decision tree, extra trees, and random forest, and trained to obtain the corresponding optimization models. Subsequently, the four models are fused to derive the optimized fusion models. The experimental results demonstrate that the fusion models generally have superior prediction performance and robustness than the traditional single-model prediction algorithms, and the analysis of the feature importance affecting grain temperature demonstrates that the factors affecting the grain temperature are, in order of importance, warehouse temperature, air temperature, warehouse humidity, and air humidity. In the future work, further study will optimize the algorithm from both temporal and spatial considerations to further improve the prediction algorithm so that it can be applied to more practical grain storage temperature monitoring scenarios.

ACKNOWLEDGMENT

This research is partly supported by Open subject of Scientific research platform in Grain Information Processing Center (KFJJ2022011), The Innovative Funds Plan of Henan University of Technology (2022ZKCJ13).